\def\l{\lambda }  \def\r{\varrho }  \def\s{$\,$}  
\def\o{\omega }  \def\a{\alpha } \def\lc{\l _{{\rm crit}}}
\def\d{{\rm d}}   \def\e{\varepsilon }
\font\tenyyy=cmcsc10 \def\yyy{\tenyyy}
\documentstyle[12pt]{article}
\newcommand{\simgeq}{\; \raisebox{-0.4ex}{\tiny$\stackrel
{{\textstyle>}}{\sim}$}\;}
\newcommand{\simleq}{\; \raisebox{-0.4ex}{\tiny$\stackrel
{{\textstyle<}}{\sim}$}\;}
\newcommand{\beq}{\begin{equation}}
\newcommand{\beqar}{\begin{eqnarray}}
\newcommand{\eeq}[1]{\label{#1} \end{equation}}
\newcommand{\eeqar}[1]{\label{#1} \end{eqnarray}}
\begin{document}
\hspace{5in}
{\sc CNLS-94-01} \par \vspace{1in}
\baselineskip 30pt
\centerline{{\huge Periodic Orbits and Shell Structure in}}
\centerline{{\huge Octupole Deformed Potentials}}
\medskip
\baselineskip 20pt
\centerline{{\yyy
W.D.\s Heiss$^{\star}$, R.G.\s Nazmitdinov$^{\star \star}$
\footnote{on leave of absence from
Joint Institute for Nuclear Research,
Bogoliubov Laboratory of Theoretical Physics, 141980 Dubna, Russia}
and S.\s Radu$^{\star}$ }}
\medskip
\centerline{{\sl
$^{\star}$ Centre for Nonlinear Studies and Department of Physics}}
\centerline{{\sl
University of the Witwatersrand, PO Wits 2050, Johannesburg, South Africa }}
\centerline{{\sl
$^{\star \star}$ Departamento de Fisica Teorica C-XI }}
\centerline{{\sl Universidad Autonoma de Madrid, E-28049, Madrid, Spain}}
\begin{abstract}
The effect of an octupole term in a quadrupole deformed single particle
potential is studied from the classical and quantum mechanical view point.
Whereas the problem is nonintegrable, the quantum mechanical spectrum
nevertheless shows some shell structure in the superdeformed prolate
case for particular, yet fairly large octupole strengths; for spherical or
oblate deformation the shell structure disappears. This result is associated
with classical periodic orbits which are found by employing the removal of
resonances method; this approximation method allows
determination of the shape
of the orbit and of the approximate octupole coupling strength for which it
occurs. The validity of the method is confirmed by
solving numerically the classical equations of motion. The quantum mechanical
shell structure is analysed using the particle number dependence of the
fluctuating part of the total energy. In accordance with the classical result
this dependence turns out to be very similar for a superdeformed prolate
potential plus octupole term and a hyperdeformed prolate potential without
octupole term. In this way the shell structure is explained at least for
some few hundred levels. The Fourier transform of the level density further
corroborates these findings.
\end{abstract}
\textheight=23cm
\newpage
\voffset=-2.2cm
\section{Introduction}
  The shell structure is one of the most important quantum features of finite
Fermi systems. The quantisation of a system of Fermions moving in a common
potential leads to a bunching of levels in the single--particle spectrum,
known as shells. When the
levels of a bunch are filled the system is particularly stable;
an additional particle fills a level in the next bunch at considerably
higher energy and therefore produces a less stable system.
In atoms and nuclei the shell effects are discerned by the
deviations of the binding energies from the smooth variation obtained from
the liquid--drop model or as oscillations in the radial density distribution
as a function of particle number and of deformation \cite{BM75}. As it was
pointed out in \cite{St68,Br72} the shells are a
global phenomena not only for
spherical systems. When a spherical shell is only
partially filled, the higher
degeneracy in the single--particle spectrum of a deformed
mean field can lead
to a breaking of spherical symmetry, and it therefore
gives rise to a deformed
equilibrium shape.

A deep understanding of shell structure phenomena in terms of classical
trajectories has been achieved by Balian and Bloch \cite{BB72} based on the
periodic orbit theory of Gutzwiller \cite{Gu71}.
According to the semiclassical
theory \cite{BB72} the frequencies in the level density oscillations of
single--particle spectra of nuclei are determined
by the corresponding periods
of classical closed orbits. The short periodic orbits give the major
contribution to the gross shell structure \cite{BM75,SM76,St77,Fr90}.
Depending on the particular mean field potential a
deviation from spherical symmetry can lead to chaotic
motion in the corresponding classical problem, and the shell structure of the
corresponding quantum spectrum is affected or even destroyed depending on
the degree of chaos \cite{Gu90,Ar87,He94}.

Atomic clusters are another area which can serve as testing ground for these
ideas. Recent experimental results on metallic clusters reporting abundance
variations in mass spectra, ionisation potentials, static polarisabilities
and collective giant dipole resonances, barrier shapes and fragmentation
provide us with striking manifestations of shell structure effects related to
a quantised motion of the valence
electrons \cite{He93,Bra93}. The correspondence
of the electronic shell structure in spherical clusters to the closing of
major quantal shells \cite{Bj90,Br93,Ma91} caused considerable interest in
using nuclear shell model type calculations for the description of metallic
clusters \cite{Cl85,Fr93,Ya93,Bu93}. It turns out that phenomenological
potentials used traditionally in nuclear physics serve a purpose similar to
those obtained within the Kohn--Sham
density--functional method \cite{KS65} if
the relevant parameters are adjusted appropriately.
Typical potentials are the
Woods--Saxon (and its various modifications) and the modified Nilsson
potential without spin--orbit term. The considerable lowering of the
computational time due to their simple analytical form renders an analysis
of the stability of large metallic clusters feasible. Naturally, the shell
numbers have to be larger than the ones used in nuclear physics in accordance
with the larger number of valence electrons considered for metallic clusters.
Recently, the semiclassical analysis has been successfully applied to explain
shell structure produced by valence electrons \cite{Ni90,Le93,Pa93}
in different metals. As it was mentioned before, obviously, shell structure
in the quantum mechanical spectrum is associated with periodic orbits in the
corresponding classical problem. Moreover, following work by Balian and Bloch
\cite{BB72}, Nishioka, Hansen and Mottelson \cite{Ni90} predicted
the supershell, which is a beating pattern, when the regular oscillations
reflecting the main shells are enveloped by an oscillating amplitude of
lower frequency. As was argued in the case of a spherical cavity \cite{BB72},
the beating pattern is caused by the interference  between  the
two most important classical trajectories which are characterised by
orbits of a triangular and square shape.
For mesoscopic objects like clusters, deformations, i.e.\s deviations
from spherical symmetry of the potential, are as important as in the nuclear
physics context (for example, the smaller peaks in the cluster abundances
correspond to unfilled shells or deformed shapes of clusters \cite{Kn85}).
Direct information about the deformation of clusters can be obtained from
an analysis of a splitting of the plasmon resonance or from a more detailed
investigation of mass abundance.

Based on a simple analysis of the geometry of closed classical orbits,
Mottelson \cite{Mot76,BM75} predicted a tendency for a large system (heavy
nuclei) with a 2:1 quadrupole deformation to exploit the octupole degree of
freedom. For nuclei this idea was supported by
Nilsson+Strutinsky shell--structure energy calculations \cite{Ab92}, by a
systematic study of octupole correlations within a microscopic approach
\cite{Eg92,Ska93} and by a
qualitative analysis of the dynamical symmetry of the anisotropic harmonic
oscillator \cite{Naz92}. Hamamoto {\it et al} \cite{Ham91} investigated the
importance of nonaxial octupole deformation for many--body system within a
harmonic oscillator model. More recently, authors in \cite{Fr93} used
a deformed Woods--Saxon potential, including quadrupole, octupole and
hexadecupole terms, to calculate the ground--state deformation of sodium
clusters.

  Inclusion of an octupole term in addition to a quadrupole term renders the
classical single particle motion nonintegrable. In fact, the system
turns out
to be chaotic. In a recent investigation \cite{Blo93} the study of classical
motion in a cavity with oscillating walls of even and odd higher order
multipoles has led to interesting conclusions
about elastic versus dissipative
behaviour of a noninteracting gas depending on the integrability or
nonintegrability of the equations of motion.

  In the present paper we thoroughly investigate a simplified
model. We leave out terms, which, albeit physically important, are prone
to blur the analysis when the interest is focussed on the essentials,
in particular a distinction between orderly and chaotic motion
in many body systems, like nuclei and metallic clusters. Since our interest
is directed not only towards the classical but also
the corresponding quantum mechanical motion, we leave out the spin--orbit
term and the $(\vec l)^2$-term present in the Nilsson model so as to render
as closely as possible the analogy between the classical and
quantum cases. An analysis of classical trajectories in a quadrupole deformed
harmonic oscillator has shown that the spin orbit term leads to chaotic
motion \cite{Roz92}. Likewise, the $(\vec l)^2$-term gives rise to chaotic
behaviour \cite{HN94} even without the octupole term. Despite and because of
the simplifications, such a model allows an understanding
of the main features of shell structure effects in super (hyper) deformed
nuclei \cite{BM75,Naz92} and in deformed clusters (see, for example,
\cite{He93,Ne92} and references quoted therein). Recent experimental
data of superdeformed K-isomers in nuclei \cite{De93} and electronic
shell structure effects in
metallic clusters \cite{He93} seem to indicate the manifestation of oblate
deformation in nuclei and small clusters. In the present paper we investigate
the effect of the octupole term for prolate and oblate deformation; the
latter case has not been dealt with in \cite{Blo93}.

One major result of the present paper is a classical analysis which
demonstrates that the prolate case including octupole deformation is
still quasi-integrable . In this way the quantum mechanical shell structure
found earlier \cite{He94} is given a proper theoretical foundation. One may
speculate that this is the reason why for large quadrupole deformation
prolate-octupole nuclei(clusters) are more stable than the oblate-octupole
ones. For metallic clusters finite temperature should be considered
\cite{Br93,Ma91}; however, our interest is focussed on shell structure whose
character is unaffected by temperature except for the amplitude. The detailed
analysis of magic numbers, their changes under variation of the octupole
strength and the possibility of supershell structure induced by the octupole
term is a further major contribution of this paper.

\section{The model}
  We investigate the classical and quantum mechanical motion in the potential
\beq
V(\r ,z)={m\over 2} \o ^2 (\r ^2+{z^2\over b^2}+\l {2z^3-3z\r ^2 \over
\sqrt{\r ^2+z^2}}).  \eeq p
For $b>1$ ($b<1$) this is a quadrupole deformed harmonic oscillator of
prolate (oblate) shape with an additional octupole term; in fact the term
proportional to $\l $ is $r^2P_3(\cos \theta )$ with $P_3$ the third
order Legendre polynomial. We use cylindrical coordinates $z$ and $\r
=\sqrt{x^2+y^2} $. For $\l \ne 0$ this is a two degrees of freedom system
which is non-integrable.

The exact classical orbits are obtained numerically by integrating the
equations of motion, {\it viz.}
\beqar \ddot \r &=&-{\partial \over \partial \r }V(\r ,z) \nonumber \\
       \ddot z&=&-{\partial \over \partial z }V(\r ,z).     \eeqar c
It is convenient to use the energy as one initial condition;
because the potential obeys the scaling law
$V(\gamma \r ,\gamma z)=\gamma ^2V(\r ,z)$ one value of the energy yields
orbits of all energies. The equation for $\r $ would contain the term
$p_{\phi }^2/\r ^3$ on the right hand side of Eq.(\ref{c}) if a finite value
of the (conserved) angular momentum $p_{\phi }$ would be considered.
However, as discussed in the following section, $p_{\phi }=0$ is sufficient
for all classical aspects relevant in this paper.

Our choice of the octupole term ensures that we have a genuine bound state
problem for $\l <\lc $. Here $\lc $ is defined to be the value for
which the potential no longer binds, and for $|\l |>\lc $ the potential
tends to $-\infty $ along one or two directions in the $\r $-$z$-plane.
The direction and the
value of $\lc $ depend on the quadrupole deformation $b$.
For prolate nuclei ($b>1$), $\lc =1/(2b^2)$ and the potential opens its
valley along the positive (negative) $z$-direction for negative
(positive) $\l $. For oblate nuclei ($b<1$) the other possible
direction along the line
$\r  ={\rm sign}(\l  )\beta z$ with $\beta \approx 0.4$,
is of increasing importance. At the value $b\approx 0.58$ valleys along the
two directions $\r  =0$ and $\r  =0.4z$
open simultaneously for $\lc \approx 1.5$ while for a still smaller
value of $b$, say for $b=0.5$, the valley along the direction $\r  =0.4z$
opens for $\lc \approx 1.64$ while the one along the $z$-axis
now opens for a larger value of $\l  $. Analytic expressions for
$\lc $ and the direction $\beta $ as functions of $b$ are given in
Appendix A.

\section{Classical perturbative treatment}
To understand the occurrence of shell structure in the corresponding
quantum mechanical problem we present a classical analysis in which it
becomes obvious that, for the prolate case, the problem is approximately
equivalent to an integrable problem even for finite values of $\l $. In
fact it is shown in what follows that, within the approximation used, the
motion in the $z$ and $\r $-direction becomes uncoupled whereby the
motion in the $z$-direction depends on $\l $. In this way frequencies in
either direction can be clearly defined with the
effect that the corresponding
winding number, i.e.\s the ratio $\o _{\r }/\o _z$, becomes a simple
function of $\l $. This provides the basic mechanism for the occurrence
of shell structures for the quantum mechanical problem; in fact it provides
classical evidence for the quantum mechanical finding that the spectrum for
specific values of $\l $ is nearly equivalent to a spectrum with $\l =0$
but with a larger value of the quadrupole parameter $b$. We stress that
the agreement between the perturbative results and the exact orbits is
excellent.

The method employed is based on the `removal of resonances' approach which
amounts to a particular averaging procedure used in secular perturbation
theory \cite{LL81}. The complete Hamilton function written in terms of the
angle and action variables of the unperturbed problem (see Appendix B) reads
\beq H(J_{\r },J_z,\a _{\r },\a _z)=\o (J_{\r }+{1\over b}J_z+\l
{\sqrt{bJ_z}\sin \a _z(2bJ_z\sin ^2\a _z-3J_{\r }\sin ^2\a _{\r })\over
\sqrt{bJ_z\sin ^2\a _z+J_{\r }\sin ^2\a _{\r}}}). \eeq h
The frequencies $\o _z ={\partial H\over \partial J_z }$
and $\o _{\r }={\partial H\over \partial J_{\r }}$ are given by
\beqar \o _z(\vec J,\vec \a )&=&{\o \over b}(1+\l
{b\sin \a _z
(4b^2J_z^2\sin ^4\a _z+6bJ_zJ_{\r }\sin ^2\a _{\r }\sin ^2\a _z-
3J_{\r }^2\sin 4 \a _{\r } )
\over 2\sqrt{bJ_z}(bJ_z\sin ^2\a _z+
J_{\r }\sin ^2 \a _{\r })^{{3\over 2}}})  \nonumber
\\
\o _{\r }(\vec J,\vec \a )&=&\o (1-\l
{\sin\a _z\sin ^2\a _{\r } \sqrt{bJ_z}
(8bJ_z\sin ^2 \a _z+3J_{\r }\sin ^2\a _{\r } )
\over 2(bJ_z\sin ^2\a _z+
J_{\r }\sin ^2 \a _{\r })^{{3\over 2}}}). \eeqar f
It is clear from these expressions that, for small values of $\l $, the ratio
$\o _{\r }/\o _z$ is essentially determined by $b$. For prolate
superdeformation, when $b=2$ or larger, one can expect that averaging
over the fast moving angle $\a _{\r }$ of the Hamilton function should yield
a good approximation. Since the Hamilton function is periodic in both angles,
this approximation amounts to keeping the zero order term of its Fourier
expansion in the fast moving angle \cite{LL81}. Since integration of the
Hamilton function over $\a _{\r }$ can be done analytically, this
approximation appears particularly attractive.

After averaging and then rewriting the action variables and the remaining
angle $\a _z$ in terms of the original momentum and coordinate values we
obtain from Eq.(\ref{h})
\beq H_{{\rm av}}={p_{\r }^2+p_z^2\over 2m}+{m\o ^2\over 2}
\biggl[\r ^2+{z^2\over b^2}+\l \xi ^2{ {\rm sign} (z)\over 2\pi }
\biggl(8{z^2\over \xi ^2}K(-{\xi ^2\over z^2})-
3\pi F_{21}({1\over 2},{3\over2},2;-{\xi ^2\over z^2})\biggr)\biggr] \eeq x
where $\xi ^2=2J_{\r }^{(0)}/(m\o )=\r ^2+p_{\r }^2/(m\o )^2$ which is
a constant within the approximation. Here $F_{21}$ and $K$ are the
hypergeometric function and the first elliptic integral, respectively.

With Eq.(\ref{x}) we have effectively reduced the full problem to an
integrable problem as the motion in the two coordinates is uncoupled. The
motion in the $z$-coordinate is determined by an effective potential which
now depends also on the action $J_{\r }$ in addition to its dependence on
$\l $. Note that the effective potential originating from the octupole term
depends only on $z/\xi ,\,\xi \sim \sqrt{J_{\r }}$;
this means that, depending
on the action residing in the $\r $-motion,  the amplitudes of the
vibrations in the $z$-motion are increased or decreased. The potential
of the $z$-motion is no longer harmonic. Because of its dependence on
$\l $ and $J_{\r }$ we now expect ratios $\o _{\r }/\o _z$ which are
different from the unperturbed value $b$ and which depend
on $\l $ and $J_{\r }$. Also note that the large $z$ behaviour coincides with
that of the complete problem which means that $\lc =1/(2b^2)$ still holds.
In Fig.(1) we illustrate the effective potential.

A few comments are in order with regard to non-zero angular momentum. The
additional term  $p_{\phi }^2/\r ^2$ in the Hamilton function does not alter
our procedure nor the results based upon it. In fact, the $\r $-motion
remains unperturbed including the centrifugal term, and for the harmonic
oscillator the frequency $\o _{\r }$ remains therefore equal to its
zero angular momentum value. This is significant in the quantum mechanical
context as the ratios $\o _{\r }/\o _z$ are independent of the angular
momentum just like in an unperturbed harmonic oscillator. It is by this
mechanism that we can explain pronounced shell structure for $\l \ne 0$,
i.e.\s shells containing all values of angular momentum which are allowed
quantum mechanically. We return to this point in the subsequent section.

The approximation procedure seems to be useful only
for the super- and hyperdeformed
prolate case. In the superdeformed oblate case (b=1/2) one would, at first
glance, consider averaging over $\a _z$. However, the zeroth order term
vanishes now as the octupole term is an odd function of $\sin \a _z$. The
first order term in the Fourier series no longer decouples the coordinates
$\r $ and $z$ so that no gain is made. For the spherical case ($b=1$) a
canonical transformation could in principle enforce one of the (transformed)
frequencies to be faster than the other. However, this procedure again is
not useful since the slow frequency turns out to be a complicated function
of time, in other words, the ratios $\o _{\r }/\o _z$ fail to be constant
rational numbers. In simple terms, this discussion shows that only the
fairly strongly deformed prolate case remains close to integrability when
the octupole
term is switched on. This is the classical explanation for the existence
of shell structure in the quantum spectrum which occurs only for the
prolate case \cite{He94}. As we see below the
spherical and the oblate case soon become chaotic classically when the
octupole is switched on, which prevents occurrence of shell structure in the
corresponding quantum spectrum.

The frequency $\o _z$ as determined by the approximation implied by
Eq.(\ref{x}) is given by $\o _z=2 \pi/T$ where
\beq T=\sqrt{2m}\int_{z_{{\rm min}}}^{z_{{\rm max}}}{\d z
\over \sqrt{E_z-U(z)}} \eeq t
with $E_z=E-E_{\r }=E-m\o ^2\xi ^2/2$ and $U(z)$ is taken from Eq.(\ref{x})
and reads
\beq U(z)=
{m\o ^2\over 2} \biggl[{z^2\over b^2}+\l \xi ^2{ {\rm sign} (z)\over 2\pi }
\biggl(8{z^2\over \xi ^2}K(-{\xi ^2\over z^2})-
3\pi F_{21}({1\over 2},{3\over2},2;-{\xi ^2\over z^2})\biggr)\biggr]. \eeq u
The period $T$ can be evaluated numerically. In Fig.(2) we display
the ratio $\o _{\r }/\o _z$  as a function of $\l $ for a few parameter
values $\xi $. Starting at its unperturbed value $b=2$ for $\l =0$, it turns
out that the ratio of the frequencies is a monotonously increasing function
of $\l $ for $0\le \l <\lc $ for a large range of $\xi $-values, for which it
is virtually independent of $\xi $; only when $\xi ^2$ approaches its upper
limit $2E/(m\o ^2)$ can a notable $\xi $-dependence be discerned. We note
that the integral in Eq.(\ref{t}) can be evaluated analytically for $\xi =0$;
in this case the ratio reads
\beq {\o _{\r }\over \o _z}=
{b\over 2}({1\over \sqrt{1+\l/\lc }}+{1\over \sqrt{1-\l/\lc }})
\eeq a
which is the top curve drawn in Fig.(2). Simple,
i.e.\s short periodic orbits occur when the frequency ratio becomes a
rational number $p/q$ with small values of $p$ and $q$. From Fig.(2) we read
off that around $\l =0.66\lc $ the ratio is 5:2 and for $\l =0.8\lc $ it is
3:1. In Figs.(3) we display the most important periodic orbits.
If the approximation is meaningful this finding implies that for these
particular values of $\l $ all orbits (except the ones with $E_z\simgeq 0$)
should be periodic with the respective winding numbers. The consequence for
the quantum spectrum is clear: the parameters $b=2$ and $\l =0.66\lc $ should
yield a spectrum similar to that for the parameters $b=5/2$ and $\l =0$;
likewise, $b=2$ and $\l =0.8\lc $ should be nearly equivalent to $b=3$ and
$\l =0$. In the following section we see a convincing confirmation of this
classical prediction of quantum shell structure.

One expects a similar pattern when starting from the outset, say, with
$b=5/2$, since the onset of classical chaos seems to be delayed
the larger the value of $b$.
In fact, one finds appropriate $\l $-values for which the
winding numbers become 3:1 (for $\l =0.63\lc$) and 7:2 (for $\l=0.76$).
Starting with $b=3$ one finds 7:2 and even 4:1, i.e.\s we have the the "chain
of shells" $2:1\to 5:2\to 3:1\to 7:2\to 4:1\ldots$ We mention here the 7:2
orbit to demonstrate the principle, but we have left out the 7:3 in the
previous paragraph although it is there between the 2:1 and 5:2. As a matter
of fact we have left out all ratios $p/q$ with $p>7$ because they do not
feature as prominently in the quantum spectrum as the shortest orbits with
$p<7$. But we can draw the conclusion that an octupole term in a prolate
deformed oscillator produces essentially a pure but more deformed
prolate oscillator. This explains the occurrence of shell structure
in the corresponding quantum spectrum at
the particular $\l $-values for which
the effective value of $b$ assumes a rational number with small value of
the numerator and denominator.

We have tested the validity of the approximation procedure by looking at a
large variety of exact orbits obtained from solving Eq.(\ref{c}) numerically.
In Fig.(4a) surfaces of section are displayed for $\l =0.63\lc $
for a number of initial conditions. The slight discrepancy between the values
0.66 found from Fig.(2) and 0.63 used in Fig.(4a) reflects the difference
between the approximation and the exact situation. Like in Fig.(2) the actual
ratio when determined numerically is an increasing function of $\l $. It
turns out that the actual ratio is slightly larger than 5:2 for
$\l =0.66\lc $ and is closest to 5:2 for $\l =0.63\lc $. What meets the eye
when looking at Fig.(4a) are the large stability islands
around the 5:2  orbit. Note that at the
periphery of the surface which is chosen to be the $\r =0$ plane we have
$\xi ^2=0$ while $\xi ^2$ attains its maximum value towards the centre. In
this way we sample the whole range. Of course we do not obtain an exact
5:2 orbit for a large range of $\xi $-values as the approximation would
suggest; we do get the one in the centre of the islands. However, while most
orbits are quasiperiodic and produce tori, their winding numbers are all very
close to 5:2, for instance 70:27 or 70:29. This applies to most of phase
space, deviations are noticeable only towards the centre of the section.
In accordance with Fig.(2) there we get orbits with smaller winding numbers
such as 7:3 and even 3:2. However, more than 85\% of phase space is dominated
by orbits with winding numbers very close to 5:2. The analogy of the
parameter set $b=2,\,\l =0.63\lc $ with $b=5/2,\,\l \approx 0$
goes even further
in that not only the shape of the trajectory of the actual 5:2 orbit is the
same but also that of the unstable orbit associated with the intersection
points of the separatrix (a separatrix similar to the one displayed in
Fig.(4a) is produced by the parameter set $b=5/2,\,\l =0.05\lc $). {\it
Mutatis mutandis} similar statements hold for the comparison of the parameter
set $b=2,\,\l/\lc  =0.76$ (see Fig.(4b)) and $b=3,\,\l \approx 0$. Note the
small region where the onset of chaotic motion can be discerned; by and large
the motion still appears close to integrabilty although $\l $ is fairly close
to $\lc $.

For comparison we display surfaces of sections in Fig.(5) for the spherical
and oblate case. Even though the diagrams represent results for smaller
values of $\l /\lc $ the overall onset of chaos is obvious. Here we note
that for the special value $b\approx 0.58$ discussed in section 2 the onset
of chaos occurs for an even smaller value of the octupole strength, i.e.\s
$\l /\lc \simgeq 0.25$.

\section{Quantum mechanical results}
The quantum mechanical treatment is straight-forward in principle. In the
spirit of previous work \cite{He} we use for the full problem which is of
the form $H_0+\l  H_1$ a representation with $H_0$ diagonal.
The basis chosen is referred \cite{BM75} to as the basis
using the asymptotic quantum numbers $n_{\perp },n_z$ and $\Lambda $ where
$n_{\perp }=n_++n_-$ and $\Lambda =n_+-n_-$. Here the quantum numbers $n_+$
and $n_-$ are the eigenvalues of $(A_+)^{\dagger }A_+$ and
$(A_-)^{\dagger }A_-$,
respectively, where, in terms of the usual boson operators $a_x$ and $a_y$,
we use $A_{\pm}=(a_x\mp i a_y)/\sqrt{2}$. For a fixed value of
$\Lambda $ (which is the quantised analogue of the classical $p_{\phi }$)
this leaves two quantum numbers (reflecting the two degrees of freedom)
to enumerate the rows and columns of the matrix problem. For $\Lambda =0$ the
diagonal entries of $H_0$ are thus $\e _{n_{\perp },n_z}^0=
\hbar \omega (n_{\perp } +1+(n_z+1/2)/b)$. The matrix elements
of $H_1$ are obtained from those of $z \sim (a^{\dagger }_z+a_z)$ and
$\r  ^2\sim (A_+(A_+)^{\dagger } +A_-(A_-)^{\dagger }+A_+A_-+(A_-)^{\dagger }
(A_+)^{\dagger })$. To get the matrix elements of $1/\sqrt{\r ^2+z^2}$ in a
numerically consistent way, we first calculate the matrix elements $S_{m,n}$
of $\r ^2+z^2$ from which the inverse square root is obtained using
$S^{-1/2}=U\cdot D^{-1/2}\cdot U^{\dagger }$ where $D=U^{\dagger }\cdot S
\cdot U$ is the diagonal form of the positive definite matrix $S$ and
$U$ is the orthogonal matrix which diagonalises $S$. To ensure also
numerically that $S$ has only positive eigenvalues it is important that
the matrix for $z^2$ is obtained by squaring $z$ and not by evaluating
analytically the matrix elements from $(a^{\dagger }_z+a_z)^2$;
inconsistencies are otherwise introduced due to truncation.
In this way, we also
ensure that the truncated matrices $S^{-1/2}$ and the representation of
$2z^3-3z\r ^2$ commute and that their product is a symmetric matrix.
The effect of truncation was tested by looking at the variation of the lower
end of the spectrum when the dimension of the matrices was increased. There
is certainly a dependence on $b$ and $\l  $. For $\l  \le 0.9\lc $
and $0.5\le b \le 2$ the variation was less than
$1 \%$ for the first 300 levels obtained from 1600$\times $1600 dimensional
matrices.

In Fig.(6a) we illustrate the spectrum so obtained as a function of $\l $ for
$b=2$. The shell structure at about $\l =0.63\lc $ and $\l =0.76\lc $
is clearly discernible. While Fig.(6a) presents the spectrum for $\Lambda =0$
we illustrate in Fig.(6b) the whole spectrum which is a superposition of all
possible $\Lambda $-values. Here the shell structure is even more pronounced
which is expected since, according to the discussion in section 3, the
orbits will have the same winding numbers
independent of the angular momentum; as a consequence, the quantum spectra
will have shell structures similar to an unperturbed oscillator. This is a
prototype case for employing Strutinsky's procedure \cite{St68,Br72} to
exhibit the magic numbers associated with the shells. However, while we
follow the basic idea of Strutinsky's method, we use a somewhat different
approach which appears more suitable in our case; this is
presently discussed.

It is clear that for the plain quadrupole deformed
oscillator, supershell structure in the sense of
\cite{Ni90} cannot occur \cite{Ar94} since there is basically only
one orbit for each coordinate; it is the sinusoidal motion characterised by
the respective frequency associated with that coordinate. The consequences
for the quantum spectrum are well known \cite{BM75}: one obtains the
Fibonacci numbers as degeneracies in the equidistant spectrum. If the
frequencies ratios are simple rationals, there are systematic repetitions of
the same degeneracies \cite{Be81,Naz92}. This is in contrast
to, say, the motion in a cavity or in a Woods-Saxon potential where basically
different classical orbits such as the triangle, the square, the pentagon
and so forth exist thus giving rise to interference of different shells in
the quantum spectrum. One of the questions addressed in this paper is the
possible occurrence of supershell effects when a quadrupole deformed harmonic
oscillator is perturbed by an anharmonic term, i.e.\s the octupole
deformation.

To analyse the quantum spectrum we proceed in the orthodox fashion in that
the total energy $E_{{\rm tot}}(N,\l )=\sum _i^N \e _i(\l )$ is approximated
by a smooth average function $S(z,\l )$ and the fluctuating difference
\beq \delta E(N,\l )=E_{{\rm tot}}(N,\l )-S(N,\l ) \eeq  d
is then subjected to further investigation. The finding
of a suitable form for
the average function $S(z,\l )$ is facilitated in our case
as it is well known
\cite{BM75} that the leading term of $E_{{\rm tot}}(N,\l )$, as a
function of
$N$, is proportional to $N^{4/3}$. We determine the five constants
$a_0(\l ),\ldots ,a_4(\l )$ in $S(z,\l )=\sum_{k=0}^4a_k(\l )z^{k/3}$ by a
least square fit which turns out to be perfectly satisfactory for all
values of $0\le \l <\lc $.

In Fig.(7) the fluctuating part $\delta E(N,\l )$ is presented as a contour
plot. The diagram refers to $b=2$ and
displays the ranges $100\le N\le 700$ and $0\le \l <\lc $.
For $\l =0$ (the bottom horizontal line) we clearly discern the shell
structure of the plain deformed oscillator.
Note that the sharp minima (dark shadowing) occur at the positions $N$
where a shell is closed; the distances are proportional to $N^3$.
When $\l $ is switched on the shell structure persists to a great extent;
only when $\l $ approaches its critical value (top horizontal line) the
structure begins to be washed out. There are local minima discernible at
$\l /\lc \approx 0.76$. This is a reflection of the enhanced shell structure
discussed in the previous section. Below we pursue this point in more detail.
The finer structure and in particular the detailed
shell structure is obtained
from the second derivative of $\delta E(N,\l )$ to which we turn next.

In Figs.(8) we have plotted
$g(E)=\delta E(N+1,\l )+\delta E(N-1,\l )-2\delta E(N,\l )$
as a function of $N^{1/3}$ for a few characteristic values of $\l $.
The peaks of the plots represent the magic numbers which characterise the
shells, and the heights of the peaks reflect the energy distance from one
shell to the next. In Fig.(8a) the essentially unperturbed result ($\l /\lc
=0.15$) is presented for demonstration. The second row displays
the results for the particular values of $\l $ for which the winding numbers
are 5:2 ($\l /\lc =0.63$) and 3:1 ($\l /\lc =0.76$), respectively. The magic
numbers and the heights of the peaks agree well with those which are obtained
from the unperturbed ($\l =0$) quadrupole deformed oscillators
(third row in Fig.(8)) with $b=5/2$ and
$b=3$, respectively, at least for $N\simleq 700,\, N^{1/3}\simleq 8.88$.
The agreement extends in particular to the respective occupation numbers,
i.e.\s the degeneracies; of course, the heights of the peaks do not show the
same regularity as the corresponding unperturbed problem; nevertheless,
even for the heights an overall agreement prevails when comparing with
Figs.(8e) and (8f) where the respective unperturbed structures are
displayed. For higher values
of $N$ we do get deviations which reflect upon the fact that the system is
nonintegrable and cannot give complete order in all its results. While the
agreement for lower values of $N$ was to be expected from the discussion in
the previous section, the extent of the agreement is rather remarkable,
especially for $\l \approx 0.76\lc $ where an astoundingly clean shell
structure reoccurs after it partially disappeared
for a somewhat smaller value
of $\l $. It is this recurrence of shell structure which gives
rise to the local minima in Fig.(7) as pointed out above.

Of particular interest is Fig.(8b) which refers to the intermediate value
$\l =0.7\lc $ where the genuinely different orbits with winding numbers
5:2 and 3:1 coexist. The long wave length fluctuation could well be
interpreted as a supershell structure, even though that it is not as clearly
pronounced as in a more transparent integrable case \cite{Ni90}. Yet the
difference to Fig.(8d) which refers to a larger value of $\l $ is striking.

In Figs.(9) we display the square of the modulus of the Fourier transform
of the level density, i.e.\s the function
\beq F(t)=|\sum_n e^{i\e _n t}|^2. \eeq y
The spectrum is taken at $\l =0.63\lc $ (Fig.(9a)) and at $\l =0.76\lc $
(Fig(9b)), both spectra refer to $\Lambda =0$ only. The pronounced peaks
can be directly associated with the periods of the classical 5:2 and 3:1
orbit, respectively, the periods obtained from Figs.(9) are in
perfect agreement with those of the corresponding classical orbits which are
found numerically by integrating Eqs.(\ref{c}). This is a beautiful
demonstration of
Gutzwiller's trace formula \cite{Gu90}. As expected the frequencies
deviate considerably from the unperturbed values, i.e.\s from the frequency
associated with $b=2,\l =0$, but also from the frequencies
associated with $b=5/2$ or $b=3$. In units of the unperturbed value ($b=2$)
we find $T_{5:2}=1.2$ and $T_{3:1}=1.4$; the values are larger than unity in
accordance with Fig.(2).

Again we stress that the high degree of order which prevails in the
superdeformed prolate case when the octupole term is turned on, does not
exist in the corresponding oblate, in fact, not even in the spherical case.
There, chaotic behaviour becomes manifest for much lesser octupole strength,
which results in a complete disappearance of shell structure in the
quantum spectrum.

\section{Summary}
The main motivation of our work is the understanding of the nature of shell
effects of many--body systems like nuclei or metallic cluster at
large quadrupole/octupole
deformation. It is the shell structure that is responsible for
the existence of superdeformed nuclei which nowadays is a very broad research
subject. Metallic clusters provide us with an additional challenge; this
topic is attracting more and more researchers who have traditionally
been working in nuclear
structure. In spite of the different character of the interaction in nuclei
(predominantly attractive forces between nucleons) and metallic clusters
(repulsive forces between valence electrons), the remarkable similarity in
the manifestation of shell structure in experimental data allows analysis
of both systems within the same quantum mechanical model.
We have investigated the shell structure produced by a quadrupole deformed
harmonic oscillator with an octupole term. Contrary to the case of spherical
potentials  our Hamiltonian is nonintegrable. Using the method based on
the 'removal of resonances' approach \cite{LL81}, we established a
connection between shell structures in the quantum mechanical spectrum
and periodic orbits in the corresponding classical problem. In the prolate
case the classical problem can effectively be reduced to an integrable set of
equations, since the motion in the two coordinates becomes uncoupled.
This is in contrast to the spherical and oblate cases, where the motion
becomes chaotic when the octupole term is switched on. In the prolate case,
at particular values of the strength parameter $\l$ the octupole
term produces a motion
which resembles to a great extent that of a pure but more enhanced prolate
oscillator. This provides the classical explanation for the existence of
quantum shell structure for prolate/octupole deformed
system within the model
\cite{He94}. At small values of $\l $ the most
important short orbit is the Lissajou figure with the shape of an
'eight' with winding number 2:1 (top of Fig.(3)).

The fluctuating part of the energy has been extracted from the quantum
mechanical single particle spectrum using our procedure which is a variant
of the method discussed in \cite{St68}. We have found
remarkable agreement between manifestations
of shell structure for the same values of $\l /\lc $ (Fig.7) and the ones
which give rise to
stability islands in the Poincar\'e surfaces
of sections relating to
the classical short orbits with winding numbers 2:1 ($\l \approx 0$), 5:2
$(\l /\lc = 0.63)$, 3:1 $(\l /\lc = 0.76)$ (Fig.(4)). Note that the orbit
closest to the triangle orbit, which is one of the most
important orbits in a cavity
\cite{BB72,Fr90,Ni90}, occurs for the quadrupole/octupole deformed
harmonic oscillator only at a large value of the octupole strength
parameter $\l$; the orbit has the winding number 3:2 and is associated
with the three islands in the centre of Fig.(4b), but it does not feature
in the quantum mechanical spectrum.
In the intermediate case $(\l /\lc = 0.7)$ when the orbits
with winding numbers 5:2 and 3:1 coexist the long wave length fluctuation
could be interpreted as a supershell structure. But due to the narrowness of
these orbits the supershell structure is not as clearly pronounced as in a
more transparent integrable case \cite{Ni90}.

In this paper we placed our emphasis on the superdeformed case which is the
most interesting situation when the octupole term is switched on. Shell
structure is destroyed for smaller values of the octupole strength for
lesser quadrupole deformation or even more so for the spherical case.
Our analysis
implies that the shell structure favours the superdeformed prolate/octupole
in contrast to the oblate/octupole case. The model
is only a crude approximation
of a realistic situation, and a mean field without consideration
of correlations (interactions between particles) may not justify a
statement as strong as this. Only future experimental data from the newly
built facility
like EUROGAM  and GAMMASPHERE can test our conclusion with regard
to nuclei. For clusters one could expect that the process of cluster
aggregation and cluster evaporation will follow valleys of the potential
energy surface, which is the combined effect of
electronic shell structure and
ionic lattice. The electronic shell effects should be rather important
at relatively small temperature and not for too large number of particles.
In the framework of our model the electronic shell structure becomes less
pronounced for $N > 700$, since higher up in the spectrum the chaotic
behaviour may well interfere with the search for shell structure \cite{HN94}.
Alternatively, if shell structure can be observed for larger numbers, our
conclusion would be that weaker deformation must prevail. More experimental
evidence is needed to assess the situation; such evidence is expected also
to shed light on the question of whether clusters provide further insight
on the fascinating question of quantum behaviour of classically chaotic
systems.

Finally we comment on the $(\vec l)^2$-term in the Nilsson model,
which plays a positive role as a phenomenological term in
nuclear physics. From our results \cite{HN94} one should infer that this
term destroys shell structure, since it produces chaos in the classical
analogous (quadrupole deformed) case, while the
experimental data indicate shells for clusters. This is different from
nuclear physics where chaos does not show, since one is not interested in
too many levels. Further investigations have
to provide clarity on this aspect. Work to this effect is in progress.
\vskip 1cm \noindent
RGN gratefully acknowledges financial support from DGICYT of Spain.

\newpage

\appendix
\section{Determination of $\beta $ and $\lc $}
To study the behaviour of the potential along the line
$z=\beta \r $ we consider the expression
$$ 2V(\r ,z)/(\r ^2m\o ^2)
=1+{\beta ^2\over b^2}+\l {2\beta ^3-3\beta \over \sqrt{1+\beta ^2}}.$$
The zero of the derivative gives the maximum/minimum of the leading
behaviour of the potential along the line $z=\beta \r $. From this we find
the relation
$$ {2\beta \over b^2}+\l ({6\beta ^2-3\over \sqrt{1+\beta ^2}}
-{\beta ^2(2\beta ^2-3)\over\sqrt{1+\beta ^2}^3})=0. $$
Solving for $\lc $ we obtain
$$\lc ={-2\beta \sqrt{1+\beta ^2}^3\over b^2(4\beta ^4+6\beta ^2-3)}.$$
Inserting this for $\l $ into the previous equation yields the slope
$\beta $ as a function of $b$ when $\l =\lc $, i.e.
$$\beta ={1\over \sqrt{8}}\sqrt{{9\over 2+b^2}-6+
{\sqrt{3(3+2b^2)(1+14b^2)}\over 2+b^2}}.$$

\section{Action and Angle Variables}
The relations between the coordinates and momenta and the actions $J_i$ and
angles $\a _i$ are given by
\begin{eqnarray*}
\r &=&\sqrt{{2J_{\r }\over m\o }}\sin \a _{\r } \\
z &=&\sqrt{{2bJ_z\over m\o }}\sin \a _z \\
p_{\r }& = & \sqrt{2m\o J_{\r }}\cos \a _{\r } \\
p_z&=& \sqrt{{2m\o J_{\r }\over b}}\cos \a _z.
\end{eqnarray*}

\newpage
\centerline{{\bf Figure captions}}
\vspace{0.5 cm}
{\bf Fig.1}
The potential $U(z)$ in units of $m\o ^2/2$ for a few values of $\l $ and
an intermediate value of $\xi $. Variation of $\xi $ produces basically
similar shapes.

\vspace{0.5 cm}
{\bf Fig.2}
The ratio $\o _{\r }/\o _z$ as a function of $\l $ for various values
of $\xi $. A genuine $\xi $-dependence is discernible only for $\xi >0.8
\xi _{{\rm max}}$.

\vspace{0.5 cm}
{\bf Fig.3}
The major orbits with winding numbers 2:1 ($\l \simgeq 0$), 5:2 ($\l /\lc
=0.63$) and 3:1 ($\l /\lc =0.76$).

\vspace{0.5 cm}
{\bf Fig.4}
Surfaces of sections ($b=2$) of four orbits for $\l= 0.63\lc $ (top) and
of five orbits for $\l =0.76\lc $ (bottom).

\vspace{0.5 cm}
{\bf Fig.5}
Sections of three orbits for $b=1$ (top) and for $b=1/2$ (bottom)
for $\l= 0.5\lc $.

\vspace{0.5 cm}
{\bf Fig.6}
Spectrum  ($b=2$) for $\Lambda =0$ (top) and all $\Lambda $ (bottom) as a
function of $\l $. The energy unit is $\hbar \o $. Note that there are
only avoided level crossings in the top figure which, however, are not
resolved.

\vspace{0.5 cm}
{\bf Fig.7}
Contours of $\delta E(N,\l )$ with $100<N<700$ as abscissa and $0<\l/\lc
<1$ as ordinate. Dark areas represent minima. Volume conservation is
taken into account in Figs.(7) and (8).

\vspace{0.5 cm}
{\bf Fig.8}
Shell structure for fixed values of $\l $ and $b=2$. The second derivative of
$\delta E(N,\l )$ is illustrated versus $N^{1/3}$
for $\l /\lc =0.15, 0.70, 0.63$ and 0.76
in (a), (b), (c) and (d), respectively. In (e) and (f) the unperturbed
($\l =0$) structure is displayed for $b=2.5$ and $b=3$, respectively.

\vspace{0.5 cm}
{\bf Fig.9}
Fourier transforms of the modulus square of the level
density for $\Lambda =0,
b=2$ for $\l /\lc =0.63$ (top) and $\l /\lc =0.76$ (bottom).

\end{document}